\documentclass[11pt,fleqn,reqno]{article}

\usepackage[parfill]{parskip}
\usepackage{amssymb, amsmath}
\usepackage{graphicx}
\usepackage{float}
\usepackage{setspace}
\usepackage[toc,page]{appendix}
\usepackage{todonotes}
\usepackage{enumerate}
\usepackage{amsthm}
\usepackage{hyperref}
\usepackage{color}
\usepackage{enumerate}
\usepackage{graphicx}
\usepackage{subcaption}
\usepackage{mathrsfs,amsmath}   
\usepackage{tikz}

\usepackage[round]{natbib}

\let\IG\iffalse

\newtheorem{proposition}{Proposition}

\newtheorem{dfn}{Definition}

\newtheorem{observation}{Observation}[section]

\newcommand{\bpf}{\begin{singlespace} \noindent {\bf Proof}:\ \ }
\newcommand{\epf}{$\;\rule{1.5mm}{3mm}$ \end{singlespace} \smallskip}

\newcommand{\bdelta}{\boldsymbol{\delta}}

\def \lbar{\underline}

\def \til[#1]{\widetilde{#1}}

\numberwithin{equation}{section}

\begin{document}

\title{Moral Hazard with Heterogeneous Beliefs
} 

\author{Martin Dumav\thanks{Universidad Carlos III de Madrid}
\and Urmee Khan\thanks{University of California Riverside}
\and Luca Rigotti\thanks{University of Pittsburgh}
}

\date{1 October 2021}

\maketitle

\begin{abstract}

We study a model of moral hazard with heterogeneous beliefs 
where each of agent's actions gives rise to a pair of probability distributions over output levels, one representing the beliefs of the agent and the other those of the principal.
The agent's relative optimism or pessimism 
dictates whether the contract is high-powered (i.e. with high variability between wage levels) or low-powered. When the agent is sufficiently more optimistic than the principal, the trade-off between risk-sharing and incentive provision may be eliminated. Using Monotone Likelihood Ratio ranking to model disagreement in the parties' beliefs, we show that incentives move in the direction of increasing disagreement. In general, the shape of the wage scheme is sensitive to the differences in beliefs. Thereby, key features of optimal incentive contracts under common beliefs do not readily generalize to the case of belief heterogeneity.

\bigskip


\bigskip


\end{abstract}

\newpage

\section{Introduction}

Since the classic work of 
\citet{holmstrom1979moral}, moral hazard theory has focused on the interplay between differences in information and differences in risk-aversion. Incentive provision and risk-sharing motives are the two fundamental forces driving contract design. When the agent's actions are observable
and contractible, efficient contracts follow from the need to insure the parties against fluctuations in payoffs, and thus crucially depend on differences in risk-aversion. When the agent's actions are not observable or contractible, optimal contracts are no longer efficient in risk sharing because they must also provide incentives for the agent to behave in a fashion desired by the principal. Thus, a key insight in the classical model of moral hazard is the importance of the trade-off between risk sharing and incentive provision. 

We extend the canonical model by allowing agent and principal to have different beliefs about the possible output levels. We study how the trade-off between incentives and efficiency is affected by heterogeneous beliefs, and examine the robustness of some of the standard conclusions of the canonical model. How is the efficiency versus incentives trade-off affected by the absence of common beliefs? Is it still the case that for unobservable actions efficiency is no longer attainable? 

To address these questions we study a natural generalization of the model in \citet{grossman1992analysis}, where each of the agent's actions is associated with a pair of probability distributions over output levels, one perceived by the agent and another by the principal. This divergence of beliefs is the only departure from the traditional model as we use the  otherwise canonical set-up: the principal is risk-neutral, the agent is risk-averse, the contract stipulates a payment from the principal to the agent,  i.e. a wage, that depends on the realized output level, and the parties' beliefs are common knowledge. Indeed, a version of our model where for \emph{all} actions the principal and the agent share common beliefs specializes to the canonical model. Beliefs can differ for many reasons; for example, principal and agent may not share a common understanding of the production technology, due to possibly having diverse experiences in the past, or having information coming from different/conflicting sources. We use this model to understand the role belief divergences play in the fundamental efficiency-incentives trade-off, and to what extent the qualitative properties of optimal contracts depend on homogeneity of beliefs.

We begin by considering the perfect information case, where the agent's action is observed by the principal and can be specified in the contract. With homogeneous beliefs, optimal risk-sharing dictates that the agent receives a constant wage. This result follows because in the absence of an incentive problem the risk-neutral party absorbs all the risk; this is a straightforward implication of the Borch rule (\citet{borch1962equilibrium}) for optimal risk-sharing with common beliefs.\footnote{Formally, efficiency implies equality of the marginal rates of substitutions across different states and when beliefs are homogeneous that equality only depends on the curvature of the agent's utility function.} With heterogeneous beliefs the Borch rule implies that a constant contract is no longer efficient because the contract must account for differences in beliefs as well as the agent's risk-aversion.\footnote{Formally, efficiency still implies equality of the marginal rates of substitutions but now probabilities typically do not drop out of that equality.} 

The first-best contract in an environment with heterogeneous beliefs has a few interesting features. First, wages are not necessarily monotone in output because they must account for the \emph{relative} difference in beliefs between the two parties; the relative ``optimism'' or ``pessimism'' over a particular state dictates whether the agent will be paid higher or lower wages in that state compared to others. For example, if the agent places higher probability on some state it could be cheaper to pay him a higher wage in that state and a lower wage in another state over which the principal has more optimistic beliefs (and such a state must exist under heterogeneity). Thus, the principal can exploit the difference in beliefs and end up with a lower expected wage bill than paying a constant wage across all states. 

A similar intuition can be invoked to understand how wages react to (small) changes in one party's beliefs: our comparative static results show that in the first-best scenario wages move in the opposite direction of the change in the principal's belief. Finally, even for the first-best it is hard to determine the entire shape of the wage function in general given that wages are sensitive to belief differences. To get a sharper characterization, we examine the case where the beliefs of the two parties over states can be ranked by the Monotone Likelihood Ratio Property (MLRP). We show that if the principal's beliefs dominates that of the agent in the monotone likelihood ratio order, then cost minimizing wages are monotone decreasing, while they are monotone increasing in the opposite scenario. 

When actions are not observable, there is an added layer of complexity to the analysis as the belief structure now has multiple dimensions of comparison. In particular, we have the beliefs of one party -the agent- over multiple actions that play into the incentive constraints, and the different beliefs of the two parties over the same action that are relevant for individual rationality constraints and the principal's objective function. In the perfect information case, it is only the second lot that matters, since there are no incentive problems. In the asymmetric information case, however, the comparison between different actions matter for incentives. These elements are reflected in the first-order conditions of the problem. These conditions have different components that depend on the beliefs of different parties: one term depends on the beliefs of the principal, while the other two depend only on the beliefs of the agent. Among the latter two, one term captures the risk-sharing role of the contract while the other captures the incentive provision part. As in the standard case, the optimal wage scheme has a risk-sharing role and an incentive provision role. Unlike the standard case, however, the incentive provision term can become irrelevant, as it may be undermined by the principal's ability to exploit the differences in beliefs to provide cheaper incentives.

Because of this more complicated trade-off between risk-sharing and incentives, to get tractability we focus primarily on the model with two actions, even though some results are more general. We show that monotonicity features of the optimal wage scheme can be driven by the difference between the beliefs of the two parties, regardless of the likelihood ratio ranking over the agent's beliefs pertaining to different actions. In the standard model, monotonicity follows from the monotone likelihood property of (homogeneous) beliefs for different actions.

Finally, we provide what we believe to be novel insights into the marginal effects of disagreement on the ``power of incentives'' (here we take the variance of wages as the ``power''). Incentives move in the direction of increasing disagreement, if we consider small perturbations in beliefs. It is worth noting that no general characterization 
of such responsiveness properties of wage functions with regards to perturbations in (homogeneous) beliefs in the canonical model is available in the existing literature. In that sense, to the best of our knowledge, our characterization that incentives move in the direction of increasing disagreement is perhaps  the first result of this kind.

The closest related paper is \citet{de2011overconfidence}, which offers a particular formulation of heterogeneous beliefs and, like ours, focuses on characterizing the shape of the optimal contract. \footnote{\citet{ostrizek2020vague} studies a similar environment as part of a dynamic principal-agent model with learning, while 
\citet{adrian2008disagreement} study a continuous-time dynamic version with particular assumptions about beliefs and utility.} The main differences are that \citet{de2011overconfidence} makes specific parametric assumptions about the probability distributions on one hand, whereas we do not impose any such restrictions, and that the notion of disagreement considered in that paper depends on the optimal contract, while it is fully exogenous in our case. While there are differences in the scope and generality of the results, there are also important parallels, as we discuss below, and some of our results are generalizations of those in \citet{de2011overconfidence}. 

As our discussion above shows, there are three main forces operating in the contracting environment with heterogeneous beliefs: risk sharing, incentive provision and the principal's ability to exploit the differences in beliefs in the form of side bets. \citet{de2011overconfidence} discusses similar ideas, namely ``incentive effects'' and ``wager effects'' (i.e. side bets). With observable actions our results are a straightforward generalization of the results in \citet{de2011overconfidence}. With unobservable actions,\citet{de2011overconfidence} provides the intuition that with enough disagreement between the agent and the principal, side bets may be exploited so that the incentive constraint does not bind. However, the measure of disagreement in \citet{de2011overconfidence}- overconfidence - includes the utility function evaluated at the optimal contract and therefore is not fully exogenous, while our notion of disagreement only depend on beliefs and thus does not depend on the details of the optimal contract. Finally, \citet{de2011overconfidence} provides comparative statics on principal's profit with respect to changes in agent's overconfidence and shows that the comparative statics can go in different directions. We show that things are in fact quite ambiguous in general as far as principal's profits, or monotonicity of wage payments, are concerned.

Our paper considers arbitrary differences between the beliefs of the principal and the agent, and focuses on the characteristics of the optimal contracts under those conditions. This is related to the literature in finance focused on the consequences of managerial overconfidence that started with 
\citet{gervais2001learning}. \citet{gervais2011overconfidence}, for example, focus on the capital budgeting decisions of an overconfident manager. This approach differs from our approach as we do not explicitly focus on the agent's overconfidence; also, we pursue a  general moral hazard setting similar to that in 
\citet{grossman1992analysis}. 

Our paper also relates to the literature on robustness of contracts under ambiguity. It turns out that our setup of a moral hazard problem with probabilistic but heterogeneous beliefs can be viewed as falling somewhere in between the standard homogeneous beliefs models and models with ambiguity as sets of probabilities. Robust contracts in the latter setting turn out to have simple forms: e.g. linear (\citet{dumav2021moral}) or step functions (\citet{lopomo2011knightian}). Interestingly, the disagreement between the parties is what drives these shapes, in particular these are the contracts that either eliminate the disagreement, or exploit it, for Pareto improvement. Disagreement also drives the shape of the contracts in our model, but the contracts themselves turn out to be more complicated and generally sensitive to the details of  the belief structure. These differences in properties of optimal contracts arise because the disagreement in our  model is exogenously fixed, whereas the disagreement in models of moral hazard with ambiguity are endogenous to the contractual form. Hence in the latter setting, the contracts themselves become the tool to shape the equilibrium level of disagreement, which is not possible in our setting given exogenous heterogeneity in beliefs. 

To summarize, we are interested in the robustness of the characterization of optimal contracts  one obtains with homogeneous beliefs. In particular, we investigate how the trade-off between risk-sharing and incentive provision play out when allowing for heterogeneity in beliefs. We show that belief heterogeneity can sometimes eliminate this trade-off altogether, thus first-best and second-best contracts may coincide. The new feature of our environment is the principal's ability to exploit belief differences and provide incentives with cheaper contracts. Another important finding is that the shape of the wage scheme is now harder to pin down and is crucially sensitive to the differences in the belief structure. Thereby many of the key results found in canonical models with homogeneous beliefs regarding monotonicity of wage schemes do not readily generalize to the model with heterogeneous beliefs.  

\section{Model}

We use the classic discrete moral hazard environment of 
\citet{grossman1992analysis}. The principal owns a technology that produces stochastic output; the agent's action determines the probability distribution of output. Formally, there are $S$ states, with typical element denoted $s$, and output is given by a vector $y \in \mathbf{R}^S$; we label states so that higher states correspond to higher output, and thus $y_1<y_2<...<y_S$. The agent chooses an action $a \in \mathbf{A}$, a finite set. The principal and agent have different beliefs about how each action influences the distribution of output. Formally, for each action taken by the agent there are two probability distributions, $\pi^P(a) \in \Delta(S)$ and $\pi^A(a) \in \Delta(S)$, representing the beliefs of the principal and the agent respectively ($\Delta(S)$ denotes the S-dimensional simplex).

The principal chooses the wages paid to the agent, denoted by the vector $w \in \mathbf{R}^S$, and keeps the difference between output and wages for herself. In keeping with the traditional moral hazard model, we assume the principal is risk-neutral while the agent is risk-averse. Formally, when the agent chooses action $a$ the principal's utility function is given by
\begin{equation*}
    U^P(y,w;a) = \sum_{s=1}^{S} \pi^P_s(a)(y_s-w_s)
\end{equation*}
while the agent's utility function is
is given by
\begin{equation*}
    U^A(w;a) = \sum_{s=1}^{S} \pi^A_s(a)u(w_s)-c(a)
\end{equation*}
where $c(a)$ represents the cost to the agent of taking action $a$. As typical in this literature, we assume the agent is risk averse so that $u'>0$ and $u'' < 0$. Finally, unless we specify otherwise, in what follows we will focus on the problem of finding the contract that implements a given action, rather than discussing the optimal action.

\section{Symmetric Information}

The first-best contract corresponds to the situation in which the agent's action is observable and the principal can explicitly make it part of the contract. In this case, the problem reduces itself to a standard risk sharing problem. When beliefs are identical, this problem has a well-known solution: pay the agent for the utility of the outside option plus the disutility of taking the desired action via a constant wage across states. This result follows from the simple observation that the (Pareto) optimal risk-sharing solution for a risk-neutral principal and risk averse agent entails the principal carrying all the risk. In our setting this result no-longer holds because (Pareto) optimal risk sharing between a risk-neutral principal and a risk-averse agent with heterogeneous beliefs will imply the latter still carries some risk.

This is easy to see by looking at the first-best problem for the principal:
\begin{align*}
    & \max_{w \in \mathbf{R}^S} \sum_{s=1}^{S} \pi^P_s(a)(y_s-w_s) \\
       & \qquad \text{subject to} \\
    & \sum_{s=1}^{S} \pi^A_s(a)u(w_s)-c(a) \geq \bar{u}
\end{align*}
which yield the following first-order condition
\begin{equation}\label{eqn:first-best-FOC}
        \pi^P_s(a)= \lambda \pi^A_s(a)u'(w_s) \qquad \forall s=1,...S
\end{equation}
where $\lambda$ is the Lagrange multiplier corresponding to the individual rationality constraint. Simple manipulation of equation (\ref{eqn:first-best-FOC}) implies that an optimal contract must solve
\begin{equation*}
    \pi^P_s(a)= \frac{\pi^A_s(a)u'(w_s)}{\sum_{s=1}^{S}\pi^A_s(a)u'(w_s)}  \qquad \forall s=1,...S
\end{equation*}
This is the standard condition for Pareto optimal risk-sharing between individuals who have different beliefs, and it can also be written as equality between marginal rates of substitutions of the agent and the principal across any two states. Denote $w^{FB}(a)$ a contract that solves the principal's problem corresponding to action $a$, and let $C^{FB}(a;\pi^P(a))$ the expected cost for the principal of that contract according to the agent's beliefs. When $\pi^P_s(a)= \pi^A_s(a)$ for all $s$, equation (\ref{eqn:first-best-FOC}) implies that $u'(w_s)$ must be constant; this in turns implies the value of the constant $w$ is found by solving individual rationality as an equality; in that case the first-best cost is $C^{FB}(a;\pi^A(a))=h(\bar{u}+c(a))= C^{FB}(a;\pi^P(a))$. If $\pi^P_s(a) \neq \pi^A_s(a)$ in at least two states, a constant payment $w$ cannot solve condition (\ref{eqn:first-best-FOC}) and therefore the first-best wages cannot be constant. In this case, the principal takes advantage of the disagreement with a contract that ends up being cheaper than $h(\bar{u}+c(a))$.We have
\begin{equation*}
    C^{FB}(a;\pi^A(a))=h(\bar{u}+c(a))>C^{FB}(a;\pi^P(a))
\end{equation*}

\begin{figure}[H]
\centering
\begin{tikzpicture}[line cap=round,line join=round]
\draw[->,ultra thick] (0,0)--(10,0) node[right]{$w_0$};
\draw[->,ultra thick] (0,0)--(0,10) node[above]{$w_1$};

\draw [line width=0.5pt,dash pattern=on 1pt off 1pt,domain=0:9.2] plot(\x,{\x});
\draw[color=black] (9.5,9.5) node {$45^{\circ}$};

\draw [line width=0.25pt,dash pattern=on 1pt off 1pt,domain=0:3] plot({\x},3);
\draw (0,3) node [left] {$h(\bar{u}+c(a))$};

\draw [line width=0.25pt,dash pattern=on 1pt off 1pt,domain=0:2.6] plot({\x},2.6);
\draw (0,2.6) node [left,color=red] {$C^{FB}(a;\pi^P(a))$};


\draw[line width=3pt, color=green,domain=0.9:10,samples=100] plot(\x,{exp ( (1/2)^(-1) * ( (1/2)   * ln (3)   + (1/2) * ln (3)   - (1/2) *   ln (\x) ) )   });


\draw [line width=2pt,color=red    ,domain=0.25:3.4]   plot(\x,{( (3/2) * sqrt(3)) /(1/4) - ( (3/4) / (1/4)) * (\x)});
\draw [line width=2pt, dash pattern=on 1pt off 3pt, color=orange ,domain=0.2:5.8]   plot(\x,{( (1/2) * 3 + (1/2) * 3 ) /(1/2) - ( (1/2) / (1/2)) * (\x)});
\draw [line width=2pt,color=blue   ,domain=0.25:8]   plot(\x,{( (2) * sqrt(2) ) /(2/3) - ( (1/3) / (2/3)) * (\x)});

\draw[color=red] (3,7.5) node [right, rectangle, rounded corners=1mm, thick, draw=black!50, fill=black!10, text width = 6cm] {First-best contract when principal is more optimistic about state 0};
\draw[color=red,<-,line width=2pt] (1.75,5.25) -- (3,7.5);

\draw[color=orange] (5.5,5) node [right, rectangle, rounded corners=1mm, thick, draw=black!50, fill=black!10, text width = 4.5cm] {First-best contract when principal and agent agree};
\draw[color=orange,<-,line width=2pt] (3,3) -- (5.5,5);

\draw[color=blue] (6,3) node [right, rectangle, rounded corners=1mm, thick, draw=black!50, fill=black!10, text width = 6cm] {First-best contract when principal is more optimistic about state 1};
\draw[color=blue,<-,line width=2pt] (4.35,2.175) -- (6,3);

\end{tikzpicture}
\caption{First-Best Contracts With Two States} \label{fig:first-best}
\end{figure}
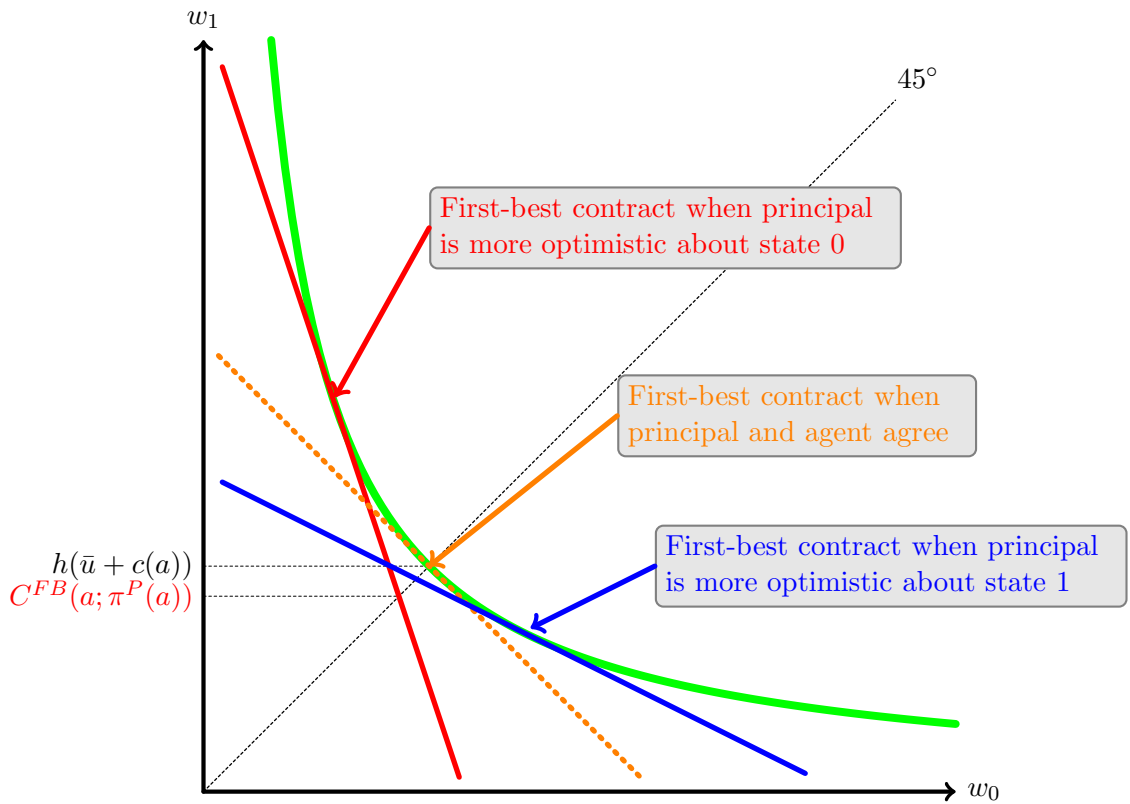

These results are easily illustrated graphically in the case of two states. In Figure \ref{fig:first-best}, the vertical axis measures the principal's payment to the agent in the state where output is high, while the horizontal one is the payment in the state where output is low; the straight lines represent the principal's expected costs under different probabilities; the agent's indifference curve corresponding to an action $a$ has the usual shape. In the dashed orange expected-costs line, the principal's beliefs are identical to the agent; in this case, the first-best contract is to pay the agent a constant wage, equivalent to Pareto efficient risk-sharing between a risk-neutral and a risk-averse individual respectively. 

As Figure \ref{fig:first-best} shows, when the principal is more optimistic than the agent about a state, the payment in that state will be lower; again, this is a simple consequence of Pareto efficient risk sharing. Since this is true irrespective of the output associated with that state, the first-best contract is not necessarily monotone in output in general. However, one can say something about monotonicity of the contracts if the beliefs of the principal and the agent are ordered based on the following incomplete order.

\begin{dfn}\label{def:MLRP}
\normalfont
Two probability distributions $f$ and $g$ over $\Delta(S)$ are ranked according to the Monotone Likelihood Ratio Property (MLRP), and $f$ is said to dominate $g$ in MLRP, if $\frac{f(s)}{g(s)}$ is increasing in $s$. In that case, we write $f \succsim^{MLRP} g$.
\end{dfn}

Since the labeling of states reflects the ordering of output levels, if the monotone likelihood ratio order holds between the principal's and the agent's beliefs for a given action, first-best contracts for that action are monotone in output. This is a rather simple result whose proof we include for the sake of completeness.

\begin{proposition}\label{prop:first-best-monoton}
\normalfont
If $\pi^P(a) \succsim^{MLRP} \pi^A(a)$ then the first-best contract is monotone decreasing in output. If $\pi^A(a) \succsim^{MLRP} \pi^P(a)$ then the first-best contract is monotone increasing in output.
\end{proposition}

\begin{proof}
The result follows from the first-order condition of the first-best problem. For any two states $s$ and $s'$ with $s>s'$, equation (\ref{eqn:first-best-FOC}) implies
\begin{equation*}
    \frac{\pi^P_s(a)}{\pi^P_{s'}(a)}=\frac{\pi^A_s(a)}{\pi^A_{s'}(a)}\frac{u'(w_s)}{u'(w_{s'})}
\end{equation*}
This expression can be rearranged as
\begin{equation*}
    \frac{\pi^P_s(a)}{\pi^A_s(a)}=\frac{\pi^P_{s'}(a)}{\pi^A_{s'}(a)}\frac{u'(w_s)}{u'(w_{s'})}
\end{equation*}
By Definition \ref{def:MLRP}, if $\pi^P(a) \succsim^{MLRP} \pi^A(a)$ then $\frac{\pi^P_s(a)}{\pi^A_s(a)} \geq \frac{\pi^P_{s'}(a)}{\pi^A_{s'}(a)}$ and therefore the first order condition can only hold if $\frac{u'(w_s)}{u'(w_{s'})} \geq 1$ which implies $w_s \leq w_{s'}$ by the concavity of $u$. The proof of the other case follows analogous reasoning.
\end{proof}

Regardless of monotonicity, one can ask how the optimal contracts responds to (possibly small) changes in the beliefs of one party. In particular, we consider perturbing the principal's beliefs so that the probabilities of only two states are different relative to the original distribution. In this case, the contract will reflect these changes in an intuitive way as stated in the following proposition.

\begin{proposition}\label{prop:first-best-compstat}
\normalfont
Consider $\pi^P(a)$ and $\til[\pi]^P(a)$ such that: (i) $\pi_t^P(a)=\til[\pi]_t^P(a)$ for all $t \neq s,s'$, (ii) $\pi_s^P(a) = \til[\pi]_s^P(a) + \varepsilon$ , and (iii) $\pi_{s'}^P(a) = \til[\pi]_{s'}^P(a) - \varepsilon$, with $\min(\pi_s^P(a),\pi_{s'}^P(a)) >\varepsilon>0$. Let $w^{FB}(a)$ and $\til[w]^{FB}(a)$ be the first-best contracts corresponding to $\pi^P(a)$ and $\til[\pi]^P(a)$ respectively. Then, $w_s^{FB}(a) \leq \til[w]_s^{FB}(a)$ and $w_{t}^{FB}(a) \geq \til[w]_{t}^{FB}(a)$ for all $t \neq s$ with at least two of the inequalities being strict.
\end{proposition}

The proposition says that if the principal gives more weight to one state (and less to some other state) the contract must reflect this by paying the agent less in that state, and more in some of the others.

\begin{proof}
The proof uses the first order conditions and the concavity of the agent's utility function. 
The first order conditions in state $s$ and any state $t \neq s $ for the two different beliefs of the principal imply
\begin{equation*}
    \frac{\til[\pi]^P_s(a)+\varepsilon}{\pi^P_{t}(a)}=\frac{\pi^A_s(a)}{\pi^A_{t}(a)}\frac{u'(w_s^{FB}(a))}{u'(w_{t}^{FB}(a))}
    \qquad \text{and} \qquad
    \frac{\til[\pi]^P_s(a)}{\til[\pi]^P_{t}(a)}=\frac{\pi^A_s(a)}{\pi^A_{t}(a)}\frac{u'(\til[w]_s^{FB}(a))}{u'(\til[w]_{t}^{FB}(a))}
\end{equation*}
Putting the two together one gets
\begin{equation*}
    \frac{\til[\pi]^P_s(a)+\varepsilon}{\pi^P_{t}(a)}=\frac{\til[\pi]^P_s(a)}{\til[\pi]^P_{t}(a)}\frac{u'(w_s^{FB}(a))}{u'(w_{t}^{FB}(a)))}\frac{u'(\til[w]_{t}^{FB}(a))}{u'(\til[w]_s^{FB}(a))}
\end{equation*}
Since $\varepsilon>0$ and $\pi^P_{t}(a) \leq \til[\pi]^P_{t}(a)$ for all $t$, this last equality can only be satisfied if
\begin{equation*}
    \frac{u'(w_s^{FB}(a))}{u'(\til[w]_s^{FB}(a)))}\frac{u'(\til[w]_{t}^{FB}(a))}{u'(w_{t}^{FB}(a))}>1
\end{equation*}
This shows that both fractions cannot be smaller than 1, and either $w_s^{FB}(a) \leq \til[w]_s^{FB}(a)$, or $w_{t}^{FB}(a) \geq \til[w]_{t}^{FB}(a)$ for all $t$, or both. Moreover, by individual rationality one cannot have that $w^{FB}(a)$ pays less than $\til[w]^{FB}(a)$ in every state, while optimality implies that it cannot pay more in every state. This establishes the result.
\end{proof}

\normalfont{Proposition}~\ref{prop:first-best-compstat} illustrates how changes in beliefs lead to changes in the first-best contract. While the proposition is written in terms of changes to the principal's beliefs, one can obtain a similar result by perturbing the beliefs of the agent.

\subsection{First-Best Contract with CARA Utility}
\label{sec:CARA-risk-sharing}
Here we illustrate the content of \normalfont{Proposition}~\ref{prop:first-best-compstat} by assuming the agent's utility function belongs to the constant relative risk-aversion class, and there are only three states.
In particular, assume the Agent's Bernoulli utility function over monetary outcomes has the following form $u(x) = - e^{-r x}$ where $r > 0$ and $x > 0$.
For a wage level $w$, the agent's utility is given by: $- e^{ - (w)}$.
Similarly, the outside option is represented
by a certain monetary outcome $\alpha$
which yields to the agent utility represented by $- e^{-\alpha}$.
Here it is worth recalling that utilities in exponential family
takes on negative values.
For simplicity, we take $r = 1$ and the state space $S = \{1,2,3\}$.

In this setting, an \textbf{interior solution} solves the FOC:
\begin{equation}
\label{eqn:CARA-FB-FOC}
\pi^P_{s}(a) = \left( \lambda \pi^A_{s}(a)  \right) e^{- (w_s(a) - c(a))} \quad s \in \mathcal{S}
\end{equation}
and IR in this setting takes the following form:
\begin{equation}
\label{eqn:CARA-FB-IR}
\sum_{s \in \mathcal{S} } \pi^A_s(a)  ( - e^{-(w_s(a) - c(a))} ) = \bar{u} + c(a)
\end{equation}

The FOC implies that
\begin{equation}
\label{eqn:CARA-FOC-pairwise}
e^{ - w_{s'}(a)} =   \frac{\pi^A_{s}(a)}{\pi^A_{s'}(a)} \frac{\pi^P_{s'}(a)}{\pi^P_{s}(a)} e^{- w_s(a)}
\end{equation}

Notice from \eqref{eqn:CARA-FOC-pairwise}
that consistent with \normalfont{Proposition}~\ref{prop:first-best-monoton}
in the optimal risk-sharing contract the wage scheme satisfies:
$w_s^{\ast}(a)$  is increasing in $s$  if  $\pi^A(a) \succsim^{MLRP} \pi^P(a)$;
$w_s^{\ast}(a)$  is decreasing in $s$  if  $\pi^P(a) \succsim^{MLRP} \pi^P$.

In particular, using \eqref{eqn:CARA-FOC-pairwise}  and taking $s = 1$
yields the following relationship
\begin{equation}
\label{eqn:FOC-iso-1}
e^{ - w_{s'}(a)} =   \frac{\pi^A_{1}(a)}{\pi^A_{s'}(a)} \frac{\pi^P_{s'}(a)}{\pi^P_{1}(a)} e^{- w_1(a)}
\end{equation}
Using \eqref{eqn:FOC-iso-1} together with IR now yields
\[
e^{- w_1^{\ast}(a)}
= - \frac{\pi^P_1(a)}{\pi^A_{1}(a)}
(\bar{u} + c)
\]
In turn, \eqref{eqn:FOC-iso-1} now yields
the  wage scheme $(w_s^{\ast})$ in the optimal
risk-sharing allocation
\begin{equation}
\label{eqn:CARA-risk-sharing-wages}
 e^{- w_s^{\ast}(a)}
= - \frac{\pi^P_s(a)}{\pi^A_{s}(a)}
(\bar{u} + c) > 0
\end{equation}
Notice also from \eqref{eqn:CARA-risk-sharing-wages}
that the optimal $w_s^{\ast}(a)$
is decreasing in $\pi^P_s(a)$
and does not depend on $\pi^A_{s'}(a)$.

To illustrate the comparative statics of \normalfont{Proposition}~\ref{prop:first-best-compstat} in this setting, consider the $\varepsilon$ re-allocation
between the probabilities of state $2$ and state $3$,
i.e., $\pi^P_2(a) + \varepsilon$ and $\pi^P_3(a) -  \varepsilon$.
One can observe from \eqref{eqn:CARA-risk-sharing-wages}
that as $\varepsilon > 0 $ increases in the optimal
risk-sharing contract:
$w_2^{\ast}(a)$ decreases,
$w_3^{\ast}(a)$ increases, while
$w_1^{\ast}(a)$ remains unchanged.

\section{Moral Hazard with Asymmetric Information}

We now move on to analyzing the second-best situation in which the principal cannot observe the agent's behavior and therefore actions are not contractible. The optimal contract is such that the agent voluntarily chooses the action the principal would like to see implemented. In order to implement the desired action, the principal now has to deal with an incentive compatibility constraint in addition to the individual rationality constraint.

In what follows, we assume that a more costly action entails an ``improvement" of both the principal's and the agent's beliefs in the sense of MLRP and Definition \ref{def:MLRP}. Formally, this means that $c(a)>c(a')$ implies $\pi^P(a)\succsim^{MLRP}\pi^P(a')$ and $\pi^P(a)\succsim^{MLRP}\pi^P(a')$.

The principal's problem is to maximize her utility subject to individual rationality and incentive compatibility. This problem can be divided into two steps: (i) for any given action, find the optimal payment schedule that incentivizes the agent to take that action, and (ii) given these payment schedules, choose the action to incentivize. As in the previous section, we focus on the first of these problems as our attention is centered around optimal incentive schemes.

In the first step the principal solves
\begin{align*}
    & \max_{w \in \mathbf{R}^S} \sum_{s=1}^{S} \pi^P_s(a)(y_s-w_s) \\
       & \qquad \text{subject to} \\
    & \sum_{s=1}^{S} \pi^A_s(a)u(w_s)-c(a) \geq \bar{u} \\
    & \sum_{s=1}^{S} \pi^A_s(a)u(w_s)-c(a) \geq \sum_{s=1}^{S} \pi^A_s(a')u(w_s)-c(a') \qquad \forall a' \in \mathbf{A}
\end{align*}

Similar to the standard case with homogenous beliefs, this problem is equivalent to minimizing expected wages, as expected revenues can be treated as a constant. Therefore, the principal's problem can be rewritten as

\begin{align*}\label{principal-wages}
    & \min_{w \in \mathbf{R}^S} \sum_{s=1}^{S} \pi^P_s(a)w_s \\
       & \qquad \text{subject to} \\
    & \sum_{s=1}^{S} \pi^A_s(a)u(w_s)-c(a) \geq \bar{u} \\
    & \sum_{s=1}^{S} \pi^A_s(a)u(w_s)-c(a) \geq \sum_{s=1}^{S} \pi^A_s(a')u(w_s)-c(a') \qquad \forall a' \in \mathbf{A}
\end{align*}

 There is another way to look at the principal's problem via a change of variables. The original problem has a linear objective function and non-linear constraints. The change of variables makes the constraints linear and the objective concave. This second method starts by defining a new variable $v_s=u(w_s)$ and a function $h(v_s)$ such that $h(v_s)=w_s$ (clearly, $h=u^{-1}$ whenever $u$ is invertible). The principal's problem can then be rewritten as
\begin{align*}
    & \min_{v \in \mathbf{R}^S} \sum_{s=1}^{S} \pi^P_s(a)h(v_s) \\
       & \qquad \text{subject to} \\
     & \sum_{s=1}^{S} \pi^A_s(a)v_s-c(a) \geq \bar{u} \\
    & \sum_{s=1}^{S} \pi^A_s(a)v_s-c(a) \geq \sum_{s=1}^{S} \pi^A_s(a')v_s-c(a') \qquad \forall a' \in \mathbf{A}
\end{align*}

Regardless of which problem one considers, standard arguments imply that the individual rationality constraint must hold as an equality; if not, the payment can be reduced by the same infinitesimal amount in all states without affecting the incentive compatibility constraints.

In what follows, we start by illustrating what happens in the two-actions and two-states case. This simple case lets us graphically illustrate properties of the optimal incentive scheme along with comparative statics.

\subsection{Two Actions and Two States}

In this section we focus on the simple two-states and two-actions case. This will enable us to illustrate some of the major issues one faces when going to a more general setting, and explain some of the intuition surrounding the model. We will use $H$ and $L$ to denote the two actions available to the agent, with $c(H)>c(L)$, and will assume that output is such that the principal always prefers to implement the high effort action $H$ unless otherwise stated.  We assume that higher costs imply a ranking of distributions according to monotone likelihood ratio, namely $\pi^P(H) \succsim^{MLRP} \pi^P(L)$ and $\pi^A(H) \succsim^{MLRP} \pi^A(L)$.

\begin{figure}[H]
\centering
\definecolor{green1}{rgb}{0,0.55,0}
\definecolor{green2}{rgb}{0,0.75,0}

\begin{tikzpicture}[line cap=round,line join=round]
\draw[->,ultra thick] (0,0)--(12,0) node[right]{$w_0$};
\draw[->,ultra thick] (0,0)--(0,10) node[above]{$w_1$};

\draw [line width=0.5pt,dash pattern=on 1pt off 1pt,domain=0:9.2] plot(\x,{\x});
\draw[color=black] (9.5,9.5) node {$45^{\circ}$};

\draw [line width=0.25pt,dash pattern=on 1pt off 1pt,domain=0:3] plot({\x},3);
\draw (0,3) node [left] {$h(\bar{u}+c(H))$};

\draw [line width=0.25pt,dash pattern=on 1pt off 1pt,domain=0:2] plot({\x},2);
\draw (0,2) node [left] {$h(\bar{u}+c(L))$};



\draw[line width=3pt, color=green1,domain=1.25:6,samples=100] plot(\x,  { exp ( (1/4)^(-1) * ( (1/4)   * ln (2)   + (3/4) * ln (2)   - (3/4) *   ln (\x) ) )   });
\draw[line width=3pt, color=green2,domain=0.105:10,samples=100] plot(\x,{ exp ( (3/4)^(-1) * ( (3/4)   * ln (3)     + (1/4) * ln (3)     - (1/4) *   ln (\x) ) )   });

\draw [line width=2pt,color=orange    ,domain=0.1:8 ]        plot(\x,  {    ( (3/4)^(-1)  *( (3/4)   *   3*sqrt(3/2)   + (1/4)   *   2*sqrt(2/3)     - (1/4)   *     (\x) ) )   });
\draw [line width=2pt,color=blue      ,domain=0.1:8 ]        plot(\x,  {    ( (7/8)^(-1)  *( (7/8)   *   3*sqrt(3/2)   + (1/8)   *   2*sqrt(2/3)     - (1/8)   *     (\x) ) )   });
\draw [line width=2pt,color=red       ,domain=0.1:2 ]        plot(\x,  {    ( (1/4)^(-1)  *( (1/4)   *     5.2         + (3/4)   *   0.574           - (3/4)   *     (\x) ) )   });


\draw[color=green1] (1.2,8.5) node [right, rectangle, rounded corners=1mm, thick, draw=black!50, fill=black!10, text width = 6.5cm] {Low effort agent's indifference curve};
\draw[color=green2] (0.1,9.5) node [right, rectangle, rounded corners=1mm, thick, draw=black!50, fill=black!10, text width = 6.5cm] {High effort agent's indifference curve};

\draw[color=red] (3,6.5) node [right, rectangle, rounded corners=1mm, thick, draw=black!50, fill=black!10, text width = 8.5cm] {Second-best and first-best contracts can coincide when principal is more optimistic about state 0};
\draw[color=red,<-,line width=2pt] (0.6,5.2) -- (3,6.5);

\draw[color=orange] (5.5,4.5) node [right, rectangle, rounded corners=1mm, thick, draw=black!50, fill=black!10, text width = 5cm] {Second-best contract when principal and agent agree};
\draw[color=orange,<-,line width=2pt] (1.75,3.75) -- (5.5,4.5);

\draw[color=blue] (5,1) node [right, rectangle, rounded corners=1mm, thick, draw=black!50, fill=black!10, text width = 6.5cm] {Second-best contract when principal is more optimistic about state 1};
\draw[color=blue,<-,line width=2pt] (1.7,3.5) -- (5,1);

\end{tikzpicture}
\caption{Second-Best Contracts With Two States} \label{fig:second-best}
\end{figure}
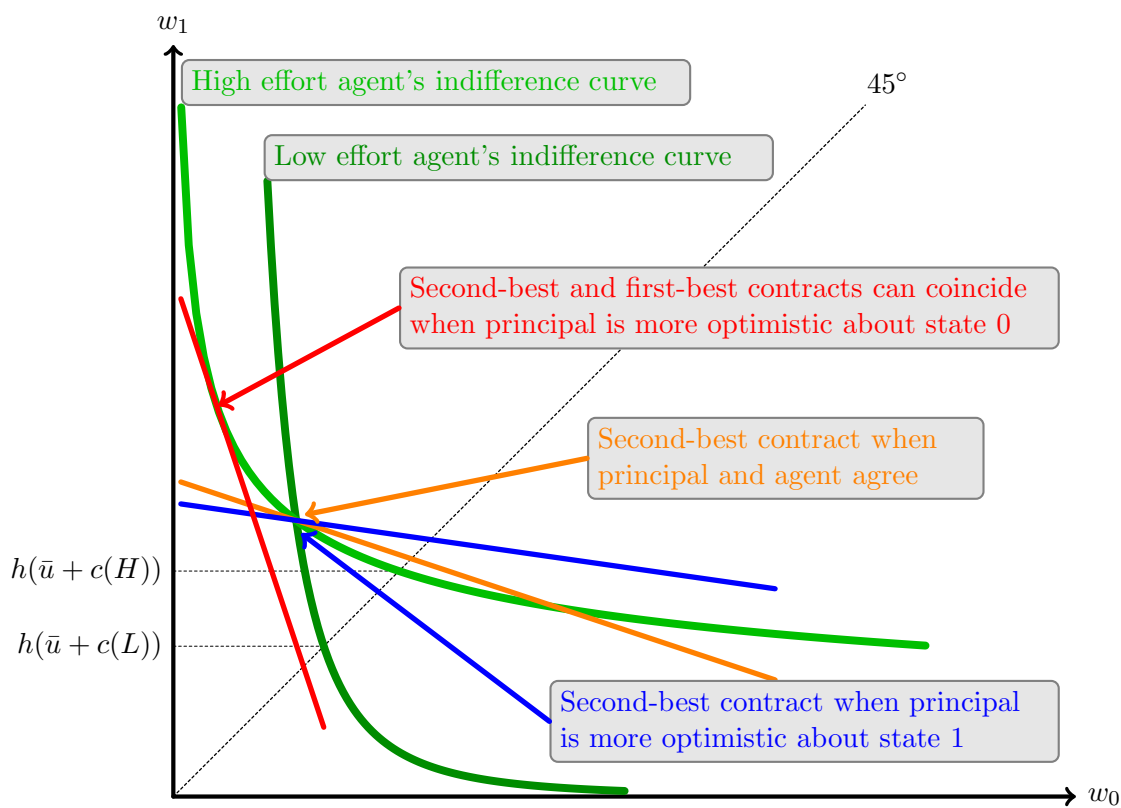

Figure \ref{fig:second-best} has payment in the high output state on the vertical axis and payment in the low output state on the horizontal one. The agent's indifference curves are those in green, the steeper darker green one corresponding to the low effort action. Each indifference curve crosses the $45^{\circ}$ line at the wage level that gives utility equal to the utility of outside option plus the cost of the corresponding effort level. The set of contracts that satisfies both constraints is given by the area above the $45^{\circ}$ line and between these indifference curves.

The principal's expected costs are given by straight lines, where lower lines imply lower expected costs and flatter lines imply higher probability of state $1$. Three of these lines are drawn, each corresponding to different beliefs. In \textcolor{orange}{orange} is the expected cost for the case in which the principal and the agent agree about the probability distribution of output when high effort is chosen. In \textcolor{blue}{blue} is the case when the principal is more optimistic than the agent about state 1, while in \textcolor{red}{red} is the case in which the principal is more optimistic than the agent about state 0.

The second-best contract is found at the lowest expected cost line that touches the area where both constraints are satisfied (that is, the area above the high effort indifference curve and below the low effort one). One can easily see that in many cases the optimum will be at the lowest `corner' of that region. In other words, with two states and two actions, in some solutions both individual rationality and incentive compatibility must bind at the optimum and the second-best contract is found at the point of intersection of the agent's indifference curves corresponding to the two actions. This is easy to see in figure \ref{fig:second-best} for the \textcolor{orange}{orange} and \textcolor{blue}{blue} expected-cost lines. In this case, the contract does not react to (small) changes in the slope of the principal's expected-cost line.\footnote{Note this would hold even in the extreme case in which the principal assigns probability $0$ to state $0$.} We summarize this first result as follows.

\begin{observation}
If the principal is not too optimistic about state $0$, the optimal contract is monotone in output, and it is not sensitive to small changes in the principal's beliefs.
\end{observation}

Figure \ref{fig:second-best} also illustrates what happens when the principal becomes `very' optimistic about state $0$ (this is represented by the steepest \textcolor{red}{red} expected cost line). In that case, incentive compatibility constraint no longer binds and second-best and first-best optimal contracts coincide. This gives our second observation.

\begin{observation}
When the principal becomes very pessimistic about state $1$, the optimal contract for the high action is the first-best contract.
\end{observation}

Although slightly more difficult to see graphically, when second and first best coincides it is possible for the principal to no longer wish to implement the high effort action. This could happen if $C^{FB}(H;\pi^P(H))>C^{FB}(L;\pi^P(L))$; this is a necessary but not sufficient condition for the expected profits to be larger under the low effort action. This gives the third observation.

\begin{observation}
When incentive compatibility is not binding, the action choice is the one that would obtain in the first best.
\end{observation}

The final point one can make inspired by the picture is about monotonicity of the principal's payoff in output. In the standard case, \citet{grossman1992analysis} show that when (homogeneous) beliefs for different actions are ordered by MLRP the principal and agent's payoffs are monotone in output. One can easily see that here the opposite could happen. As the principal becomes more optimistic about state $0$, the optimal contract will pay the agent more in state $1$, and thus it is possible that the principal's income in that state becomes lower than that in state $0$. This is our last observation for the two-states and two-actions case.

\begin{observation}
The principal's payoff need not be monotone in output.
\end{observation}

Non-sensitivity of the optimal contract to small changes in the principal's beliefs is peculiar to the two-states case; in this case if both constraints bind the contract is fully specified. This would not be true with more than two states. The other results, however, are more general and illustrate the many difficulties of solving for the optimal contract and/or doing comparative static in a more general model with a finite number of states.

\subsection{Two Actions and $S$ States}

Again, we focus on the cost minimization problem formulation. The principal's problem gives rise to the following first-order condition
\begin{equation}\label{eqn:second-best-wages-FOC}
        \pi^P_s(H)= \lambda \pi^A_s(H)u'(w_s) + \mu \left[\pi^A_s(H)-\pi^A_s(L)\right]u'(w_s)  \qquad \forall s=1,...S
\end{equation}

where $\lambda$ and $\mu$ are the Lagrange multipliers corresponding to individual rationality and incentive compatibility constraints respectively. Comparing with equation (\ref{eqn:first-best-FOC}) one notices that, as usual, the first-best optimality condition is modified by the presence of an extra term that depends on incentives. Interestingly, this extra term only depends of the agent's beliefs; this observation is lost when one looks at the case of homogeneous beliefs. Furthermore, the principal's beliefs corresponding to the low effort action have no impact on the optimal contract that implements  high effort  (although they matter for the choice of which action to implement). The multipliers are strictly positive when the corresponding constraint binds; while $\lambda$ cannot be zero because the individual rationality constraint must bind, the two states discussion above has already illustrated that $\mu$ might be equal to zero at an optimum. This is a major difference with the case of homogeneous beliefs.

%



As in the case of the first-best, here we are interested in establishing conditions for the contract to be monotone in output (i.e. increasing in $s$). The conditions are a more restrictive version of the ones we saw before. 

%

The set of optimal second-best contracts is characterized as the set of solutions to
\begin{enumerate}
\item the first-order conditions:
\begin{equation}\label{eqn:FOC-2byS}
     1 = \left( 
     \lambda \frac{\pi^A_s(H)}{\pi^P_s(H)}
     +
     \mu \frac{\left[\pi^A_s(H)-\pi^A_s(L)\right]}{\pi^P_s(H)} 
     \right) u'(w_s)
   \qquad \forall s=1,...S
   \tag{FOC}
\end{equation}
\item Incentive Compatibility
\begin{equation}\label{eqn:IC-2byS}
\sum_{s} \pi^A_s(H) u(w_s) - c \geq \sum_{s} \pi^A_s(L) u(w_s)
\tag{IC}
\end{equation}
\item Individual Rationality
\begin{equation}\label{eqn:IR-2byS}
\sum_{s} \pi^A_s(H) u(w_s) - c \geq \lbar{u}
\tag{IR}
\end{equation}
\end{enumerate}
Notice the conditions \eqref{eqn:FOC-2byS}, \eqref{eqn:IC-2byS}, and
\eqref{eqn:IR-2byS} characterizing
the set of optimal contracts do not depend on
the probability distribution
$\pi^P_s(L)$ of the Low action perceived by the principal.

We next turn to characterizing the optimal contract
under difference scenarios consistent with the monotone likelihood ratio ranking. There are two possibilities depending on how the parties rank the high action:
(1) $\pi^A(H) \succsim^{MLRP} \pi^P(H)$ and
(2) $\pi^P(H) \succsim^{MLRP} \pi^A(H)$.

\begin{proposition}\label{prop:MH-monoton}
\normalfont
(1): If $\pi^A(H) \succsim^{MLRP} \pi^P(H)$ then the wage scheme in the optimal contract is monotone increasing in output.
(2): If $\pi^P(H) \succsim^{MLRP} \pi^A(H)$ then the wage scheme
in the optimal contract can be non-monotone.
\end{proposition}

\begin{proof}

(1) $\pi^A(H) \succsim^{MLRP} \pi^P(H)$
implies that likelihood ratios $\frac{\pi^A_s(H)}{\pi^P_s(H)}$
and $\frac{\left[\pi^A_s(H)-\pi^A_s(L)\right]}{\pi^P_s(H)}$
are both increasing in $s$.
Since $u$ is strictly concave and $\lambda >0$ and $\mu \geq 0$,
\eqref{eqn:FOC-2byS} implies that $w_s$ is increasing in $s$.

(2) $\pi^P(H) \succsim^{MLRP} \pi^A(H)$
implies that likelihood ratio $\frac{\pi^A_s(H)}{\pi^P_s(H)}$
is decreasing in $s$. The second term on the RHS of \eqref{eqn:FOC-2byS},
$\frac{\left[\pi^A_s(H)-\pi^A_s(L)\right]}{\pi^P_s(H)}$,
is not necessarily monotone decreasing. Therefore, whether the contract is monotone in this case is ambiguous.
\end{proof}

Part (1) extends the monotonicity result of
\citet{grossman1992analysis}
under the monotone likelihood ration ranking to heterogeneous priors, provided the agent is more optimistic than the principal about higher states given that the high action is chosen. Part (2), instead, suggests that a failure of monotonicity is possible even if both parties find that the High action dominates the low action in the MLRP sense. Without more specific assumptions about functional forms of the utility functions and about the probability distributions of the parties under the two actions, a sharper characterization of the optimal contract in (2) is not generally possible. We provide examples below to illustrate various interesting cases.

Regardless of monotonicity, one can ask how the optimal contracts responds to (possibly small) changes in the beliefs of one party. Similar to what is done in Proposition \ref{prop:first-best-compstat}, we consider changing the principal's beliefs to a nearby distribution in which the probabilities of only two states are different relative to the original distribution.  One may expect that the contract will reflect these changes. For tractability, we use a parametric setting for the next result. 


\subsection{Comparative Statics in Moral Hazard-- 2-by-3 Examples with CARA utilities}
\label{Subsec:2by3CARA}

In the spirit of Proposition \ref{prop:first-best-compstat}, we analyze how the optimal contract changes as disagreement between the principal and the agent increases infinitesimally. Formally, the probability the principal assigns to some state increases by some $\varepsilon>0$, simultaneously decreasing the probability of a different state by that same amount.

For simplicity, we consider an interior optimal contract that is a unique solution to the set of first-order conditions \eqref{eqn:FOC-2byS} when both \eqref{eqn:IC-2byS} and \eqref{eqn:IR-2byS} bind.
The solution to this system of equations depends implicitly on
the parameter $\varepsilon$; formally, the  compensation scheme $(w_s^{\ast} (\varepsilon))_{\{s \in S\}}$ and the Langrange multipliers $\lambda^{\ast}(\varepsilon)$ and $\mu^{\ast}(\varepsilon)$.
Notice from \eqref{eqn:FOC-2byS} that the comparative static analysis on
the optimal incentive scheme $(w_s^{\ast} (\varepsilon))_{\{s \in S\}}$
depends on how the multipliers $\lambda^{\ast}(\varepsilon)$ and $\mu^{\ast}(\varepsilon)$ change as $\varepsilon$ increases. Even for optimal monotone contracts, i.e,
$(v_s^{\ast} (\varepsilon))_{\{s \in S\}}$ increasing in $s$ for any $\varepsilon > 0$,
$\lambda^{\ast}(\varepsilon)$ and $\mu^{\ast}(\varepsilon)$ can be
non-monotone in $\varepsilon$. This is the reason a tractable characterization of the comparative statics is not generally available without making further assumptions on the contracting environment.

Therefore, for what follows we focus on a special case
where the outcome space has $3$ elements ($\mathcal{S} = \{1,2,3\}$) and the agent's utility function belongs to the CARA family. In this family, there is a linear relationship between utility levels and the corresponding marginal utilities.  Thanks to this linearity, the system of equations that determine the optimal contract is in turn linear in marginal utilities, and can be solved explicitly. We let the agent's utility function over monetary outcomes have the form $u(x) = - e^{-x}$ where $x > 0$, and determine the optimal contract algebraically so that we can conduct the comparative statics analysis as $\varepsilon$ changes.

\begin{proposition}[Incentives move in the direction of increasing disagreement]
\label{prop:CARA-compStat}
\normalfont
Assume the principal's probability distribution is such that one can write the probabilities of states $s$ and $s'$ as $\pi^P_{s}(H) + \varepsilon$ and $\pi^P_{s'}(H) - \varepsilon$, with $\varepsilon>0$. Then, ${w_{s}}$ is decreasing in $\varepsilon$ and ${w_{s'}}$ is increasing in $\varepsilon$.
\end{proposition}

 \begin{proof}

In this setting, for an \textbf{interior solution} $w_s > 0$ the first-order conditions take the following form:
\begin{equation}
\label{eqn:CARA-FOC}
\pi^P_s(H) = \left( \lambda \pi^A_s(H) + \mu (\pi^A_s(H) - \pi^A_s(L)) \right) e^{-  w_s  },
\quad s \in \mathcal{S}.
\end{equation}
The Individual Rationality constraint is
\begin{equation}
\label{eqn:CARA-IR}
\sum_{s \in \mathcal{S}}  \pi^A_s(L) ( - e^{-w_s}  )= \bar{u} + c
\end{equation}
and the Incentive Compatibility
\begin{equation}
\label{eqn:CARA-IC}
\sum_{s \in \mathcal{S}} ( \pi^A_s(H) - \pi^A_s(L))  (-e^{-w_s}) = c
\end{equation}

Notice that the system of equations \eqref{eqn:CARA-FOC}-\eqref{eqn:CARA-IC}
is linear in marginal utilities $e^{-w}$ and in the Lagrange multipliers, $\lambda$ and $\mu$.
Using this system of equations we characterize the solution for the incentive scheme
algebraically. It will be useful to simplify
the notation and define the difference $\Delta_s := (\pi^A_s(H) - \pi^A_s(L))$.

We start by expressing the Lagrange multipliers
in terms of the wages. Summing over the FOCs yields
\begin{equation}
\label{eqn:lambda-1}
 \lambda = \pi^P_1(H) e^{w_{1}} + \pi^P_2(H) e^{w_{2}} + \pi^P_3(H) e^{w_{3}}
\end{equation}
and
using the FOC for $s=1$ we can write $\mu$ as follows:
\begin{equation}\label{eqn:mu(lambda)}
\mu =  \frac{\pi^P_1(H)}{ \Delta_1 } e^{w_{1}}
- \frac{ \lambda \pi^A_1(H)}{ \Delta_1 }
\end{equation}

Here $\Delta_1 < 0$, since $\pi^A(H) \succsim^{MLRP} \pi^A(L)$ implies
that $\frac{\pi^A_s(H)}{\pi^A_s(L)}$ is increasing in $s$.
For otherwise, $\pi^A(H)$ and $\pi^A(L)$ would not be different distributions.
For the same reason, $\Delta_3 > 0$.%
\footnote{
\label{fn:MLRP-3}
More specially,
the relationship $\pi^A(H) \succsim^{MLRP} \pi^A(L)$ implies
\[
\frac{\pi^A_1(H)}{\pi^{A}_1(L)} \leq \frac{\pi^A_2(H)}{\pi^A_2(L)} \leq \frac{\pi^A_3(H)}{\pi^A_3(L)}
\]
and hence $\frac{\pi^A_1(H)}{\pi^A_1(L)} < 1$ and
$\frac{\pi^A_3(H)}{\pi^A_3(L)} > 1$.
For otherwise, if $\pi^A_1(L) \leq \pi^A_1(H)$,
then  $\pi^A_{2}(L) \leq \pi^A_2(H)$ and $\pi^A_3(L) \leq \pi^A_3(H)$.
But this then implies that  $\pi^A(H) = \pi^A(L)$. Therefore,
$\Delta_1 < 0$ and $\Delta_3 > 0$.
}

To simplify notation in what follows, let $\kappa_{s',s} :=
{(\Delta_{s'} \pi^A_s(H) - \Delta_{s} \pi^A_{s'}(H))}
=  - \pi^A_{s'}(L) \pi^A_s(H) + \pi^A_s(L) \pi^A_{s'}(H) > 0$
whenever $s' > s$.
Now we use the linear system of equations \eqref{eqn:CARA-IR} and \eqref{eqn:CARA-IC} to express marginal utility at ${w_2}$ in terms of that at ${w_1}$ by removing the terms with $w_3$.
\begin{equation}\label{eqn:CARA-1/2}
\kappa_{3,1}  e^{-w{_1}} + \kappa_{3,2} e^{-w{_2}}
=
(- \Delta_3 \bar{u} + c \pi_3(L)) > 0
\tag{1/2}
\end{equation}
which can be solved to yield
\begin{equation}\label{eqn:w2(w1)}
e^{w_2} = \frac{\kappa_{3,2}}{ (- \Delta_3 \bar{u} + c \pi_3(L))   - {\kappa_{3,1}}e^{-w_1} } > 0
\end{equation}

Similarly, using \eqref{eqn:CARA-IR} and \eqref{eqn:CARA-IC} we express marginal utility at ${w_3}$ in terms of that at ${w_1}$:
\begin{equation}\label{eqn:CARA-1/3}
\kappa_{2,1}  e^{-w{_1}}
-
\kappa_{3,2} e^{-w{_3}}
= (- \Delta_2 \bar{u} + c \pi_2(L))
\tag{1/3}
\end{equation}
and solving
\begin{equation}\label{eqn:w3(w1)}
e^{w_3}
=
\frac{\kappa_{3,2}}{ {\kappa_{2,1}} e^{-w_1} - (- \Delta_2 \bar{u} + c \pi_2(L)) } > 0
\end{equation}

Now we use an equation that has not been used so far to solve for the term ${w_1}$.
For instance, we use the first-order condition \eqref{eqn:CARA-FOC} for $s=2$:
\[
\pi^P_2(H) = (\lambda  \pi^A_2(H) + \mu  \Delta_2) e^{- w_2}
\]
Using \eqref{eqn:mu(lambda)} the latter implies
\[
\pi^P_2(H) =
\left(
\lambda  \left( \pi^A_2(H) - \frac{  \pi^A_1(H)  \Delta_2 }{ \Delta_1 } \right)
+ \left(  \frac{\pi^P_1(H)   e^{w_1} \Delta_2 }{ \Delta_1 } \right)
\right)
e^{-w_2}
\]
Now, using  \eqref{eqn:lambda-1}
to substitute for $\lambda e^{c}$
in the last equation implies
\[
\pi^P_2(H) =
\bigg(
\left(\pi^P_1(H) e^{w_1} + \pi^P_2(H) e^{w_2} + \pi^P_3(H) e^{w_3}\right)
\underbrace{\left( \pi^A_2(H) - \frac{  \pi^A_1(H)  \Delta_2 }{ \Delta_1 } \right)}_{:= \gamma_2}
+ \left(  \frac{\pi^P_1(H)   e^{w_1} \Delta_2 }{ \Delta_1 } \right)
\bigg)
e^{-w_2}
\]

Simplifying and rearranging the previous equation implies
\[
\pi^P_2(H) (1-{\gamma_2} )=
\left(
\pi^P_1(H) e^{w_{1}} \left( \gamma_{2} + \frac{\Delta_2 }{ \Delta_1 } \right)
+
\pi^P_3(H) \gamma_{2}{e^{w_3}}
\right)
{e^{-w_2}}
\]
Finally, using \eqref{eqn:w2(w1)}
and \eqref{eqn:w3(w1)} and substituting
for $e^{w_2}$ and $e^{w_3}$, respectively, yields:
\begin{equation}
\label{eqn:CARA-Soln-w2}
\begin{split}
\pi^P_2(H) (1-{\gamma_2} )=
\left(
\pi^P_1(H) e^{w_{1}} \left( \gamma_{2} + \frac{\Delta_2 }{ \Delta_1 } \right)
+
\pi^P_3(H) \gamma_{2}
\frac{\kappa_{3,2}}{ {\kappa_{2,1}} e^{-w_1} - (- \Delta_2 \bar{u} + c \pi_2(L)) }
\right) \\
\left(
\frac{ (- \Delta_3 \bar{u} + c \pi_3(L))   - {\kappa_{3,1}}e^{-w_1} }
{\kappa_{3,2}}
\right)
\end{split}
\tag{$\ast$}
\end{equation}
Here
$\gamma_2 =
\left( \pi^A_2(H) - \frac{  \pi^A_1(H)  \Delta_2 }{ \Delta_1 } \right) = - \kappa_{2,1}/\Delta_1 > 0$
and $\gamma_2 \leq \pi^A_2(H)  \leq 1 $.
Moreover, the MLRP relationship $\pi^A(H) \succsim^{MLRP} \pi^A(L) $ implies that
$\left(\gamma_{2} + \frac{\Delta_2 }{ \Delta_1 } \right) > 0$ by an argument similar to that in the footnote \ref{fn:MLRP-3}.

The equation~\eqref{eqn:CARA-Soln-w2} (implicitly) determines
the unique solution for $w_1^{\ast}$ and relates it to the primitives in this setting.
In turn, substituting for $w_1^{\ast}$ in \eqref{eqn:CARA-1/2} and \eqref{eqn:CARA-1/3},
respectively yields the unique solutions for $w_2^{\ast}$ and $w_3^{\ast}$. Using \eqref{eqn:CARA-Soln-w2} we can show that
${w_1}$ increases in $\varepsilon$.
Suppose to the contrary that ${w_1}$ and hence $e^{w_1}$ decreases as $\pi^P_2(H)$ increases.
This implies that the left-hand side in \eqref{eqn:CARA-Soln-w2} increases.

On the right-hand side, notice that the two terms in the product are both increasing in $e^{w_1}$ (In the first term, larger $e^{w_1}$ makes the denominator a smaller positive number and hence makes the ratio larger). Under the hypothesis, both the first and the second term on the right-hand side decrease. This in turn implies that starting with \eqref{eqn:CARA-Soln-w2} holding with equality increasing $\varepsilon >0$ leads to a decrease in the RHS while the LHS increase, which contradicts optimality. This therefore shows that the optimal $w_1^{\ast}$ decreases in $\varepsilon$.

Now using the fact that $w_1^{\ast}$ increases in $\varepsilon$
implies, by \eqref{eqn:CARA-1/2} and \eqref{eqn:CARA-1/3}, that the optimal $w_2^{\ast}$ decreases and $w_3^{\ast}$ increases in $\varepsilon$, respectively. This proves the claim in the statement of Proposition for states $s = 2$ and $s' = 3$. The comparative static result for any pair of states in $\mathcal{S}$ follows using the equation~\eqref{eqn:CARA-Soln-w2} and analogous reasoning.
\end{proof}

\textbf{Remark 1:} This result contrasts with the first-best comparative statics in the same environment. In that problem, we have seen in Section~\ref{sec:CARA-risk-sharing} that similar comparative statics analysis has that the incentives move in the direction of increasing disagreement. However, in that problem the optimal first-best contract is not effected in the state whose probability is unchaged ($w_1^{FB}$), while here it is effected and in particular it increases. Thus the presence of moral hazard is reflected not only on the shape of the optimal contract, but also on the way it is affected by (small) changes in the beliefs of the principal.

\textbf{Remark 2:} Proposition \ref{prop:CARA-compStat} can be extended to include the full set of states in $\mathcal{S}$, we focus on two states only for simplicity. This also highlights how the incentives
for a state that is not included in the $\varepsilon$ reallocation can be affected
due to the moral hazard problem. Contrast this with the risk-sharing effect.
In particular, notice that the same comparative statics exercise
in the risk-sharing problem without an incentive compatibility condition
in Subsection~\ref{sec:CARA-risk-sharing} shows that
in the optimal risk-sharing contract $w_1^{FB}$ does not respond to the $\varepsilon$ re-allocation between the states $s=2$ and $s=3$.
 In contrast, in the optimal moral-hazard contract $w_1^{\ast}$ do respond to this change, and increases. This difference is due to moral hazard effect.

\subsection{On going beyond 2-action and 3-outcome setting}

Tractable general comparative statics results analogous to Proposition~\ref{prop:CARA-compStat} are difficult to obtain even in a simple two actions setting. As the principal's distribution for the High action $\pi^{P}_{s}(H)$ changes, one can see from \eqref{eqn:FOC-2byS} that the characterization of comparative static change in the optimal wage scheme $\{w_s\}$ depends on how both Lagrange multipliers behave with respect to this change.

In the 2-by-3 semi-parametric example analyzed above, we can explicitly characterize the solution for the optimal wage scheme in terms of the distributions perceived by the parties. Howeve, as the number of states in the contracting problem increases beyond $3$, algebraic characterization of the optimal contract is not tractable.

Instead of solving algebraically for the optimal contract one can consider implementing an iterative method to study its properties. In this iterative approach, we solve for the optimal contract as we iteratively increase the number of states in the problem. We can show that this iterative method finds the solution to the optimal contract. However, it does not yield a tractable enough characterization of the optimal contract that could enable us to perform comparative statics analysis. We summarize our findings below, and leave to the Appendix the details of this iterative approach to illustrate the extant difficulties in performing general comparative statics analysis beyond $3$ output levels.

We divide the cost minimization problem of implementing the High action into two steps.
The first part, referred to as `inner'  minimization program, fixes an arbitrary wage  level for the highest output level, $w_4$, and finds the least costly incentive scheme by minimizing over the wages over the first three output levels, $\{w_1, w_2,w_3\}$. It turns out that such a reduction in the outcome space by one preserves the MLRP rankings between High and Low actions. Therefore, our solution for 2-by-3 case above supplies the solution to the inner minimization program given a wage payment $w_4$ for the highest output realization.

In the second part, given the solution of the inner minimization program we solve for the remaining wage level $w_4$. We refer to this as the `outer' minimization program. The outer problem identifies the solution using the cost function from the inner minimization problem. We show in the Appendix that this iterative method yields the same solution as
that one obtains by solving the full minimization program over the four output levels.

Notice that in this iterative method, the outer minimization program uses the cost function in the inner minimization program. In turn the comparative statics analysis in this iterative method involves characterizing the behavior of Lagrange multipliers in the inner minimization problem with respect to changes in the distributions perceived by the parties. In general, we do not know how to characterize these Lagrange multipliers behavior with respect to comparative static changes. We leave it to future research to make progress on such a characterization.

\section{Conclusions}

To what extent do the classic results that characterize optimal contracts in moral hazard settings (e.g. \citet{grossman1992analysis}) hold up in an environment where the principal and the agent have probabilistic disagreements? We show that allowing for belief heterogeneity opens the door to new trade-offs that can lead to contracts with significantly different properties. Even in the symmetric information case, first-best risk sharing do not result in constant wages across states, and contracts may not even be monotone. In the asymmetric information case, incentive constraints may not always bind and, with enough disagreement between the two parties, the principal may be able to exploit the difference in beliefs and reach the first best contract that turns out to be cheaper from her perspective.

We also provide a novel set of results that look at the comparative statics properties of optimal contracts when beliefs change. Small changes in principal's beliefs have different effects on the shape of the optimal contract between the first and second best settings. In particular, the incentive effects in the second best setting amplifies the force of belief perturbations. As a result, perturbed beliefs in the second best case may alter the shape of the contract quite significantly. 

There are limitations to our analysis. First, we limit ourselves to only two actions. We also limit ourselves to finitely many states, and for some results we are restricted to only three states. What can be said about the properties of optimal contracts with arbitrary numbers of states and actions with reasonable degree of clarity and tractability is still very much an open question. 



\newpage

\appendix

\section{Appendix: An iterative approach to Moral Hazard problems}

We approach to analyze the contracting problem with four outcome levels
as one addition of new outcome level on the problem with with three outcome levels.
More specifically,
Consider the new outcome space $S' = \{1,2,3,4\}$
so that it expands the outcome space in previous section by
a new outcome level: $S' = S \cup \{4\}$.

To simplify notation  we re-define probability distributions
over $3$ outcome levels without superscripts:
$\pi = \pi^{A}(H)$, $\delta = \pi^{P}(H)$ and $\eta = \pi(L)$.
Moreover, we denote the distributions with $4$ outcomes with a $'$:
$\pi'$, $\eta'$ and $\delta'$. Analogous to \textbf{Assumption A} 
we assume that the ranking of the distributions over $4$ outcome levels satisfy:
\[
\pi' \succsim^{MLRP} \delta'  \succsim^{MLRP} \eta'
\]

When there are four outcome levels, the principal's cost
minimization programme to implement the High action
takes the following form:
\begin{equation}\label{eqn:CostMin-4}
\min_{\{ w_1,w_2,w_3, w_4: IR, IC \}}
\delta'_1 w_1 + \delta'_2 w_2 + \delta'_3 w_3 +  \delta'_4 w_4
\end{equation}

In this setting, we define a `reduction' of distributions
from $4$ outcomes to $3$ outcomes by `lumping' together the probabilities
on the outcome levels $s=3$ and $s = 4$ into $s= 3$. 
More specifically,
given a probability distribution over $\mathcal{S}'$
we define the corresponding `reduced' probability distributions over $3$ outcome levels as follows:
\begin{equation}
\label{eqn:Redux4-3}
\pi =
(\pi_{1} , \pi_{2}, \pi_{3})
:=
(\pi'_{1} , \pi'_{2}, \pi'_{3} + \pi'_{4})
\end{equation}
And similarly, we define
\[
\eta =
(\eta_{1} , \eta_{2}, \eta_{3})
:=
(\eta'_{1} , \eta'_{2}, \eta'_{3} + \eta'_{4})
\]
\[
\delta =
(\delta_{1} , \delta_{2}, \delta_{3})
:=
(\delta'_{1} , \delta'_{2}, \delta'_{3} + \delta'_{4})
\]

\begin{observation}
\label{obs:MLRP-redux}
\normalfont
MLRP ranking between distributions is preserved under the reduction defined in \eqref{eqn:Redux4-3}:
\[
\pi' \succsim^{MLRP} \delta'  \succsim^{MLRP} \eta'
\Rightarrow
\pi \succsim^{MLRP} \delta  \succsim^{MLRP} \eta
\]
This holds because
for any two distributions such that  $\pi' \succsim_{ML} \delta'$
and for any $s < s' < s''$
\[
\frac{\pi'_{s}}{\delta'_{s}}
\leq
\frac{\pi'_{s'} + \pi'_{s''}}{\delta'_{s'} + \delta'_{s''}}
\leq
\frac{\pi'_{s''}}{\delta'_{s''}}
\]
\end{observation}

To analyze the programme \eqref{eqn:CostMin-4}, it is without loss
to consider a change of variables involved in the principal's
cost minimization programme. The change of variable is defined by 
\begin{equation}\label{eqn:spread(m)}
m := u(w_4) - u(w_3)
\end{equation}
Here the spread $m$ in the utility space is the difference between the promised utilities
to the agent in the event of outcome realizations $s=4$ and $s=3$.
For this spread, the principal makes an additional payment $M(m)$ payment over $w_3$
if the outcome level $s=4$ realizes, i.e., $w_4 = w_3 + M(m)$. The required payment
difference $M(m) = w_4 - w_3$ implicitly defined by
\begin{equation}\label{eqn:M(m)}
m = u(w_3 + M(m)) - u(w_3)
\end{equation}
Given this change of variables, 
without loss the principal's cost
minimization programme to implement the High action
takes the following form:
\begin{equation}\label{eqn:CostMin-4Full}
\min_{\{w_1,w_2,w_3,m: IR_m, IC_m \}}
\delta_1 w_1 + \delta_2 w_2 +  \delta_3 w_3 +  \delta'_4 M(m)
\end{equation}
where
\begin{equation}
\label{eqn:IR_m}
(IR_m): \quad
\sum_{s \in S} \pi_s u(w_s) + \pi'_4 m = u_{0} + c
\end{equation}
and
\begin{equation}
\label{eqn:IC_m}
(IC_m): \quad
\sum_{s \in S} (\pi_s - \eta_s) u(w_s) + ( \pi'_4 - \eta'_4 ) m = c
\end{equation}

\subsection{Moral Hazard Problem with 4 outcomes - full programme}
We start with characterizing the solution
to the minimization programme \eqref{eqn:CostMin-4Full}
by solving for the choice variables all simultaneously.
For this, it will be useful to make the following observation:
\begin{observation}
\label{obs:M-convexity}
\normalfont
The function $M(m)$ defined implicitly by 
\eqref{eqn:M(m)}
is convex in the spread $m$.
\end{observation}

\begin{proof}
Using the definition of $M(m)$ in \eqref{eqn:M(m)}  and differentiating
with respect to $m$ yields
\begin{equation}\label{eqn:Mprime(m)}
u'(w_3 + M(m)) M'(m) = 1
\end{equation}
To show that the function $M$ is convex, consider an increase in the spread $m$.
Since $M$ is increasing in $m$ and utility function $u$ is concave, i.e., $u'' < 0$,
the first term on the LHS goes down, to maintain the equality then
$M'(m)$ must increase in $m$. The latter implies that $M''(m) > 0$ and hence
$M$ is a convex function of $m$.
\end{proof}

Since the objective function in \eqref{eqn:CostMin-4Full}
is convex, the first-order conditions
are sufficient and necessary to characterize the optimal contract.
The optimal contract $(w_s^{\ast})$ and
the Lagrange multipliers at the optimum $\lambda^{\ast}$
and $\mu^{\ast}$ solve the first-order conditions 
\begin{equation}
\label{eqn:FOCm(s)}
\delta_{s} = \left( \lambda \pi_{s} + \mu (\pi_{s} - \eta_{s}) \right) u'(w_s)
\quad s = 1,2,3
\end{equation}
\begin{equation}
\label{eqn:FOCm(m)}
\delta'_{4} M'(m) = \left( \lambda \pi'_{4} + \mu (\pi'_{4} - \eta'_{4}) \right)
\end{equation}
together with Individual Rationality $(IR_m)$ and Incentive Compatibility $(IC_m)$.

\subsection{Moral Hazard Problem with 4 outcomes - iterative programme}

We now approach to solve the optimization
problem \eqref{eqn:CostMin-4Full} in two steps.
In particular, these two steps involve a double minimization problem:
\begin{equation}\label{eqn:Double-Min}
\min_{m}
\delta'_4 M(m) +
\left(
\min_{ \{w_1,w_2,w_3: IR_{m}, IC_{m} \} }
\delta_1 w_1 + \delta_2 w_2 + \delta_3  w_3
\right)
\end{equation}
For a given spread $m$ we denote the cost function
in the inner minimization problem by
\begin{equation}
\label{eqn:Inner-Min}
C(m;\bdelta) =
\min_{\{w_1,w_2,w_3: IR_{m}, IC_{m} \} }
\delta_1 w_1 + \delta_2 w_2 + \delta_3 w_3
\end{equation}
In this inner minimization programme, the spread level $m$
 and the probability distribution
$\bdelta = (\delta_1,\delta_2,\delta_3)$ are parameters. For a given $m$ it is a cost minimization
programme that solves for $3$ wage levels for the first three output levels: $(w_s)_{s \in \mathcal{S}}$.
Its solution is given by the analysis with $3$ outcome levels as in Subsection~\ref{Subsec:2by3CARA}.
In particular, the solution to the inner minimization problem for a given spread level $m$
is characterized by
\begin{equation}\label{eqn:FOC-3m}
FOC(w_s): \quad
\delta_{s} = \left( \lambda \pi_{s} + \mu (\pi_{s} - \eta_{s}) \right)u'(w_s)
\quad s = 1,2,3
\end{equation}
\begin{equation}\label{eqn:IR-3m}
IR_{m}: \quad
\sum_{s=1,2,3} \pi_s u(w_s) + \pi_4 m = \bar{u} + c
\end{equation}
\begin{equation}\label{eqn:IC-3m}
IC_{m}: \quad
\sum_{s=1,2,3} (\pi_s - \eta_s ) u(w_s) - (\pi_4 - \eta_4) m= c
\end{equation}
Denote by $(w_s(m;\bdelta))_{s \in \mathcal{S}}$ the solution
for wage payments
and by $\lambda(m;\bdelta)$ and $\mu(m;\bdelta)$
Lagrange multipliers. 
Dependence of
the solution to the inner minimization programme
on the given spread $m$ and the distribution $\bdelta$
is noted.

Given the solution to the inner minimization
the iterative programme \eqref{eqn:Double-Min} then
reduces to determine the optimal $m$ which solves the outer
minimization programme
\begin{equation}
\label{eqn:min-iterative}
\min_{m}
\delta'_4 M(m) + C(m;\bdelta)
\end{equation}

\begin{observation}
\normalfont
The objective function in the programme \eqref{eqn:min-iterative} is convex in $m$.
\end{observation}
\begin{proof}
We have seen that the function $M(m)$ is convex in $m$.
It suffices therefore to show that the cost function
$C(m;\bdelta)$ of the inner minimization programme \eqref{eqn:Inner-Min}
is convex in $m$ for any $\bdelta$.

Notice that the constraint set \eqref{eqn:IR_m}
and \eqref{eqn:IC_m} is linear in the spread $m$ and
the promised utility levels $v_s = u(w_s)$.
Fix  a $\bdelta$ and Consider for $m_1$ and $m_2$ the convex mix
$m = \alpha m_1 + (1 - \alpha ) m_2$ with $\alpha \in (0,1)$.
By linearity, the promised utilities defined by
$\tilde{v}_s = \alpha(v_s^{\ast} (m_1)) + (1 - \alpha) (v_s^{\ast} (m_2))$
for $s \in \mathcal{S}$ satisfy  both \eqref{eqn:IR_m} and \eqref{eqn:IC_m}
and therefore feasible given the spread $m$.
The optimal incentive scheme $(v_s^{\ast} (m))$ for the required
spread $m$ achieves a weakly smaller cost relative to
the incentive scheme $\{\tilde{v}_s\}_{s \in \mathcal{S}}$
and hence $C(m;\bdelta) \leq \alpha C(m_1;\bdelta) + (1- \alpha) C(m_2;\bdelta)$,
showing that the cost function $C(m;\bdelta)$ is convex in the spread $m$.
\end{proof}

Since the objective function in \eqref{eqn:min-iterative}
is convex, sufficient and necessary first-order conditions on $m$:
\[
FOC(m): \quad
\delta'_ 4 M'(m) + C'(m;\bdelta) = 0
\]

Notice that the spread $m$ is a parameter
in the inner minimization programme \eqref{eqn:Inner-Min}.
By Envelope Theorem on the inner minimization
programme
\begin{equation}
\label{eqn:Cprime(m)}
C'(m;\bdelta) = -( \lambda(m;\bdelta) \pi'_4 + \mu(m;\bdelta) \Delta'_4)  < 0
\end{equation}
where $\lambda(m) > 0$ and $\mu(m) \geq 0$
are the Lagrange multiplies on $IR_m$ and $IC_m$, respectively,
and $\Delta'_4 > 0$ since $\pi' \succsim_{MLRP} \eta'$ (by an argument analogous to 
that made in the footnote \ref{fn:MLRP-3}).
Intuitively higher spread $m$ relaxes both constraints $IR_{m}$ and $IC_{m}$
in the inner minimization programme and 
this helps the principal to save on the expected wage payments.
In turn, the solution to the outer minimization problem is characterized by
\begin{equation}\label{eqn:FOC(m)}
\delta'_4 M'(m) = - C'(m;\bdelta)
\quad \rightarrow \quad
\delta'_4 M'(m) =  ( \lambda(m;\bdelta) \pi_4 + \mu(m;\bdelta) \Delta_4)
\end{equation}
Denote by $m^{\ast}$ the optimal spread that solves the outer minimization problem.

Finally, turning the inner minimization problem and substituting this optimal spread $m^{\ast}$
for $m$ in \eqref{eqn:FOC-3m}, \eqref{eqn:IR-3m} and \eqref{eqn:IC-3m}
the solution to the inner minimization problem
is given by $(w_1(m^{\ast};\bdelta),w_2(m^{\ast};\bdelta),w_3(m^{\ast};\bdelta))$ for the wage payments
together with the Lagrange multipliers $\lambda(m^{\ast};\bdelta)$ and $\mu(m^{\ast};\bdelta))$.
This therefore concludes the two-step procedure 
and characterizes the solution to the double minimization programme
\eqref{eqn:Double-Min}.

\subsection{Equivalence between the solution of 
the full programme and
the solution of the iterative programme}

The equivalence between the solutions follow from noticing that
the set of conditions
\eqref{eqn:FOCm(s)}, \eqref{eqn:FOCm(m)},
\eqref{eqn:IR_m} and \eqref{eqn:IC_m}
that solves full optimization problem \eqref{eqn:CostMin-4Full}
is identical to the set of equations
\eqref{eqn:FOC-3m}, \eqref{eqn:FOC(m)},
\eqref{eqn:IR-3m} and \eqref{eqn:IC-3m}
that solves the optimization programme iteratively in two step
in \eqref{eqn:min-iterative}.
It is therefore without loss to employ the iterative approach with a double minimization approach to solve for the optimal contract.
In particular, the optimal wage scheme satisfies
$w^{\ast}_s = w_s(m^{\ast};\bdelta)$ for $s \in \mathcal{S}'$
and the Lagrange multipliers satisfy $\lambda^{\ast} = \lambda(m^{\ast};\bdelta)$ and
$\mu^{\ast} = \mu(m^{\ast};\bdelta)$.

\subsection{On comparative statics with iterative approach}
We next turn to investigate 
whether using the two-step procedure to cost mininimization programme
provides a tractable way to analyze $\varepsilon$ reallocation analogous to 
Proposition~\ref{prop:CARA-compStat}
when there are four outcomes.
This turns that without a characterization of 
the behavior of Lagrange multipliers in the inner minimization 
programme tractable comparative statics results 
are difficult to obtain in general.
We illustrate this in what follows.

For comparative static analysis, we 
express the conditions that 
characterize the solution in terms 
of the primitives in the environment. 
Consider first the outer minimization programme.
Using \eqref{eqn:M(m)} and \eqref{eqn:Cprime(m)}
the first-order condition  \eqref{eqn:FOC(m)} that 
determines the optimal $m$ is given by
\begin{equation}
\label{eqn:CARA4-CompM}
\delta'_4 M'(m) = \delta'_4 \frac{1}{u'(w_3( m; \bdelta) + M(m))}
=  ( \lambda(m;\bdelta)\pi'_4 + \mu(m;\bdelta) \Delta'_4)
\end{equation}
We next consider the solution to the inner minimization 
programme. For any given $m$ the inner minimization 
involves three wage levels. Its solution therefore 
obtains in an analogous way to the equation $(\ast)$
that characterizes the solution to the optimal contract
for the case with three outcome levels. 
The solution to the inner-minimization programme $w_s(m;\bdelta)$
is characterized
by
\begin{equation}
\label{eqn:Mstar}
\begin{split}
\delta_2 (1-{\gamma_2} )=
\left(
\delta_1 e^{w_{1}} \left( \gamma_{2} + \frac{\Delta_2 }{ \Delta_1 } \right)
+
\delta_3 \gamma_{2}
\frac{\kappa_{3,2}}{ {\kappa_{2,1}} e^{-w_1} - ((- \Delta_2 \bar{u} + c \eta_2) + \kappa_{4,2} m ) }
\right)\\
*
\left(
\frac{ ((- \Delta_3 \bar{u} + c \eta_3) + \kappa_{4,3} m)  - {\kappa_{3,1}}e^{-w_1} }
{\kappa_{3,2}}
\right)
\end{split}
\tag{M*}
\end{equation}

In the reallocation of $\varepsilon$ between any 
pair of elements in $(\bdelta,\delta_4) = (\delta_1,\delta_2,\delta_3,\delta_4)$,
the comparative statics is performed using \emph{both} \eqref{eqn:CARA4-CompM} and \eqref{eqn:Mstar}.
In the analysis when there are three outcome case
as in Subsection~\ref{Subsec:2by3CARA}, 
we implicitly substituted out the Lagrange
multiplier and hence the comparative statics analysis is tractable.
When there are more than three output levels, 
this is not generally applicable.
In the four outcome case, for instance, without knowing
the behavior of Lagrange multipliers in \eqref{eqn:CARA4-CompM}
with respect to changes in $\bdelta$,
we are not yet able to perform comparative statics in a tractable manner
using the iterative method described here.
One can of course solve algebraically and perform
the comparative static analysis.
However, such an algebraic approach is not typically tractable
even in special cases, for instance, working with CARA utility functions.

\newpage

\bibliographystyle{chicago}
\bibliography{MhHet}
\end{document}